\documentclass{aastex631}

\usepackage{graphicx}
\usepackage{amsmath}
\usepackage{amsfonts}
\usepackage{bm}
\usepackage{hyperref}

\begin{document}

\title{Coupled-mode theory for astrophotonics}

\author[0000-0001-8542-3317]{Jonathan Lin}

\affiliation{University of California Los Angeles \\
Physics \& Astronomy Department \\
475 Portola Plaza \\
Los Angeles, CA 90095, USA}



\begin{abstract}
Coupled-mode theory (CMT) is a powerful tool for simulating near-harmonic systems. In telecommunications, variations of the theory have been used extensively to study waveguides, both analytically and through numerical modelling. Analogous mathematical techniques to the CMT are also widely used in quantum mechanics. The purpose of this work is to collect different formulations of the CMT and their underlying connections to quantum mechanical techniques, and to showcase their utility in modelling slowly varying waveguides including directional couplers and photonic lanterns. My choice of example waveguides is motivated by the astronomical applications of such devices in starlight nulling, wavefront sensing, and high-resolution spectroscopy. I first provide a brief review of the standard form of the CMT, applicable for waveguides with fixed eigenmodes. Next, I show that the CMT also applies for slowly varying waveguides, and demonstrate the close relation between the CMT and several well-known approximation methods from quantum mechanics, as well as concepts like geometric phase. Finally, I present a verification of my analysis, in the form of the numerical package \texttt{cbeam}.

\end{abstract}



\section{Background}

\subsection{Astrophotonics}
The cross-disciplinary field of astrophotonics applies photonic devices to the technical challenges of astronomy \citep{roadmap}, and has since led to novel demonstrations of photonic nullers \citep{glint,Xin}, spectrometers \citep{pradip,Lin:21}, wavefront sensors \citep{Norris:20,Lin:23}, and wavefront correctors \citep{diab}. Such devices offer new ways to process the light collected by astronomical telescopes, in a form factor much smaller than bulk-optical components. Figure \ref{fig:intro} (panels a \& b) show two examples of astrophotonic devices: the directional coupler \citep{AgrawalDC}, which acts as a photonic beamsplitter, and the photonic lantern (PL; \citealt{Leon-Saval,Birks}), which converts multi-moded light into single-moded light. With recent improvements in the fabrication capability for such devices, the astrophotonics community is now poised to consider how to tune designs for better performance in astronomical applications. This might include optimizing directional couplers or tricouplers \citep{Klinner_Teo_2022} and phase shifters for broadband nulling, or designing PLs which simultaneously maximize wavefront sensing and spectroscopy capability.
\\\\
To do so requires a physical understanding of how light propagates through astrophotonic devices. At least within the astrophotonics community, this understanding is often heavily reliant on numerical modelling techniques such as the finite-difference beam propagation method, which are accurate but opaque and difficult to intuit. At the same time, design optimizations for photonic devices would benefit from simulation methods that are faster than the standard methods used today.
\\\\
The goal of this paper is to introduce a mathematical tool that can be used to model and understand the propagation of light through a ``slowly-varying'' photonic device, through the lens of astronomical instrumentation. The precise meaning of the slowly-varying constraint will be developed later in this work, but for now it suffices to say that this constraint applies to the vast majority of astrophotonic devices, including PLs, photonic integrated circuits, arrayed waveguide gratings, directional couplers, multimode interferometers, fiber Bragg gratings, and Kerr combs; and that photonic devices not satisfying the slowly-varying constraint should probably be redesigned, if possible, to satisfy this constraint and simplify the modelling process. The mathematical tools reviewed in this work all fall under the broad category of ``coupled-mode theory'' (CMT). 

\subsection{Introduction to coupled-mode theory}
\begin{figure}
    \centering
    \includegraphics[width=\linewidth]{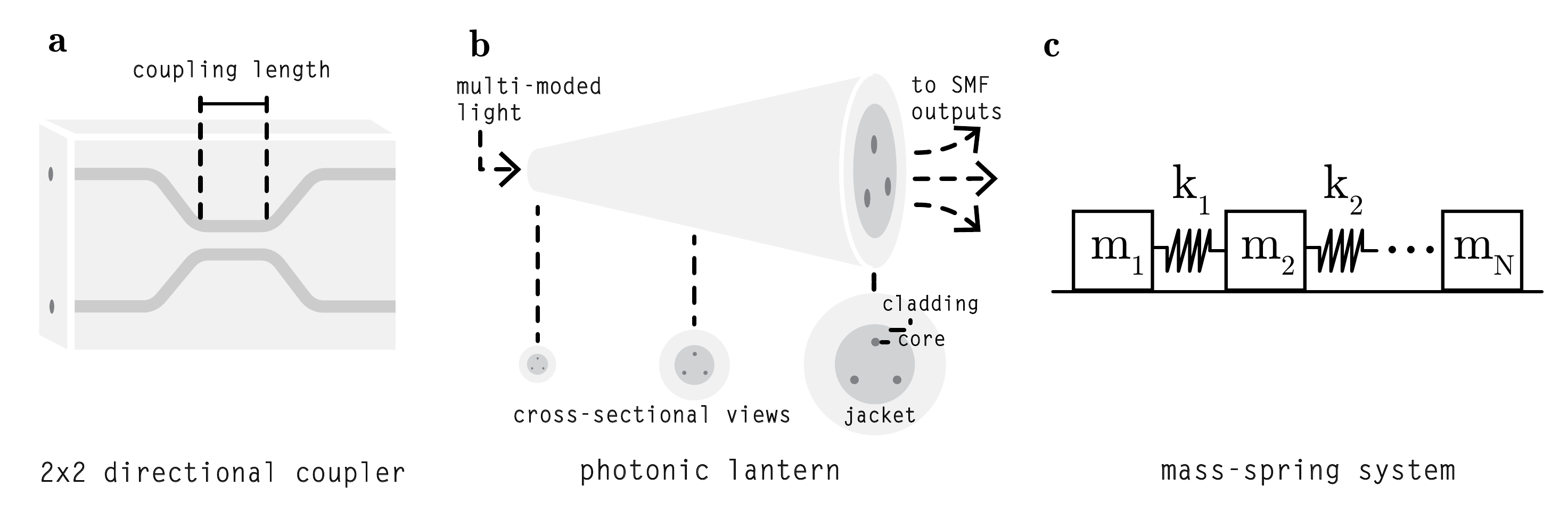}
    \caption{\textbf{a.} A $2\times 2$ directional coupler, formed from two single-moded waveguide channels which brought close enough to transfer power and then separated again. \textbf{b.} A photonic lantern: an optical waveguide which looks like a few-moded optical fiber on one end and splits off into multiple single-moded outputs at the other. The propagation of light through this device also can be modeled as coupled harmonic oscillators. \textbf{c.} A mechanical system of coupled harmonic oscillators composed of masses and springs, constrained to move in one dimension along a frictionless surface. }
    \label{fig:intro}
\end{figure}
The CMT is a well-known and well-used mathematical tool. The origins of the CMT are often attributed to studies of radio transmission lines and electron beams by Miller and Pierce, respectively \citep{Miller,Pierce}; but the underlying ideas appear even earlier in quantum mechanics, for instance in studies of molecular dynamics by Born and Oppenheimer \citep{BornOpp}. In telecommunications, the CMT has been widely used to study the propagation of light through few-moded optical waveguides, both analytically and numerically \citep{okamoto,AgrawalDC,Haus}. 
\\\\
To motivate the CMT, we will first consider a mechanical analog. Suppose we have a system of $N$ masses connected by springs, as shown in Figure \ref{fig:intro}c; our goal is to determine the dynamics of the system given an arbitrary initial condition. We first consider the dynamics of a {\it single} mass-spring system (where the other end of the spring is fixed), which is a harmonic oscillator obeying
\begin{equation}
    \dfrac{d^2 x}{dt^2} = -\dfrac{k}{m}x 
\end{equation}
where $k$ is the spring constant, $m$ the mass, and $x$ denotes position. The corresponding solution for $x(t)$ is the complex exponential. If we chain many mass-spring systems together, the dynamics of the system is that of a many harmonic oscillators, which can also transfer energy between each other. Mathematically, the equation of motion is 
\begin{equation}\label{eq:ms}
    \dfrac{d^2 \bm{x}}{dt^2} = -A \bm{x}
\end{equation}
where $\bm{x} = [x_1,x_2,...,x_N]^T$ are the positions of each mass and $A$ is an $N\times N$ matrix whose off-diagonal terms correspond to the linkage between the individual mass-spring systems. To solve this equation, we assume the solution is harmonic in time, i.e. we propose the ansatz $\bm{x} \propto \bm{u} e^{\i\omega t}$. This reduces equation \ref{eq:ms} to the eigenvalue problem
\begin{equation}
    \omega^2 \bm{u} = A\bm{u}
\end{equation}
where $\bm{u}$ and $\omega^2$ are identified as an eigenvector-eigenvalue pair. The motion of all the masses in the system may therefore be expanded in terms of the eigenvectors, whose amplitudes oscillate harmonically in time; this is ``normal mode analysis''.
\\\\
So far, our analysis has been exact. We now consider the case where the the matrix $A$ is allowed to vary with time, for instance through a gradual weakening of the springs or application of an external force. This is the problem that coupled-mode theory (CMT) attempts to solve. The principle of the CMT is as follows: suppose there is a system which supports a basis of harmonic oscillator eigenmodes. When a slowly varying perturbation is added, we assume that perturbed solutions will be ``almost'' harmonic, and attempt to find solutions formed from a linear combination of the unperturbed eigenmodes. To account for the perturbation, the modes are allowed to cross-couple through varying mode coefficients --- hence the name ``coupled-mode'' theory. 
\\\\
Existing treatments in the context of optical waveguides typically provide the ``fixed-basis'' formulation of the theory, suitable for optical fibers or composite waveguides composed of individual channels, such as directional couplers \citep{AgrawalDC}, but not for slowly varying waveguides whose eigenmodes change incrementally with longitudinal coordinate $z$. A motivating example for such a waveguide is the PL, which in recent years has been the subject of growing interest for astronomical applications such as telluric line suppression \citep{Trinh:13}, high-resolution spectroscopy \citep{Jovanovic:17,Lin:21}, starlight nulling \citep{glint,Xin}, wavefront sensing \citep{Lin:23}, and spectroastrometry \citep{Kim}. 
\\\\
The main purpose of this paper, comprising \S\ref{ssec:deriv}-\ref{ssec:deriv2}, is to provide an overview for the fixed-basis CMT as well as for the more broadly applicable ``varying-basis'' formulation \citep{SnyderLove:1983}, in some instances called ``coupled local mode theory'' (CLMT) \citep{Jin:10,chen:20}, and to demonstrate its applicability to slowly varying waveguides. Unlike some previous implementations \citep{chen:20,sunder} which assume non-degenerate modes, the formulation presented here is applicable even in the presence of degeneracy, and accounts for non-adiabatic cross-coupling. The secondary purpose, comprising \S\ref{sec:qm}, is to show how many of the ideas in the varying-basis CMT have well-studied counterparts in quantum mechanics: concepts including the WKB approximation, perturbation theory, the adiabatic approximation, and geometric phase. While individual connections have been often drawn in previous studies, this work aims at establishing a broader connection, hopefully introducing the reader to techniques and ideas that have normally not been associated with coupled-mode theory and which could be useful in future developments of astrophotonic devices. Throughout, derivations are written using mathematical notation borrowed from quantum mechanics; I briefly review this notation for readers who may be less familiar with it. Finally, in \S\ref{sec:prax} I verify my analysis with \texttt{cbeam}, a simple numerical implementation of the varying-basis CMT, and use it to simulate a two-channel directional coupler and a 6-port PL.

\section{Coupled-mode theory for $z$-invariant waveguides}\label{ssec:deriv}
This section presents a brief derivation of a form of the CMT using a fixed eigenbasis \citep{AgrawalMMF}, which is similar to time-dependent perturbation theory from quantum mechanics \citep{CH}. For simplicity I will also assume weak guidance, though the CMT is equally applicable to vector eigenmodes (\citealt{okamoto}; \S\ref{ssec:geophase}). First, I will introduce the necessary notation, including Dirac's bra-ket notation. Broadly speaking, the electric field within a waveguide is some complex-valued function $u(x,y,z)$, where $(x,y)$ are the transverse coordinates (perpendicular to the axis of the waveguide) and $z$ is the longitudinal coordinate (parallel to the axis of the waveguide). I will denote an inner product over the transverse directions with bra-ket notation:
\begin{equation}
    \langle u|v\rangle \equiv \iint \Bar{u} v \,dx dy. 
\end{equation}
Here, $\Bar{u}$ is the complex conjugate of $u$, and the brackets around the ``ket'' $|v\rangle$ indicate that the quantity is a vector in a complex vector space.\footnote{Specifically, an element of the Hilbert space of square-integrable functions, $L^2$. This is because we must have $\langle v|v \rangle$ finite, which is loosely the same as saying that the electric field is transversely ``bound'' by the waveguide.} The ``bra'' $\langle u|$ is a ``dual vector''; the inner product $\langle u | v\rangle$ is often referred to as the ``overlap integral'' between $|u\rangle$ and $|v\rangle$.
\\\\
Wave propagation under the approximation of weak guidance obeys the Helmholtz equation \citep{AgrawalMMF}: a wavefront function $\psi(x,y,z)$ or equivalently a wavefront ket $|\psi\rangle$ obeys
\begin{equation}
     \left[\nabla^2  + k^2n_0^2\right] | \psi\rangle = 0
\end{equation}
where $k$ is the free space wavenumber and $n_0=n_0(x,y)$ is the refractive index profile. Since the waveguide structure does not change with $z$, solutions should also not change with $z$ (modulo a phase factor), leading to the ansatz
\begin{equation}
    |\psi\rangle = |\phi_j\rangle e^{i\beta_j z}
\end{equation}
where $\beta_j$ is some constant of propagation, and $|\phi_j\rangle$ denotes a square-integrable function $\phi_j(x,y)$, which represents the electric field of a guided mode: a particular distribution of light which does not change shape as it propagates down the waveguide. Substituting this ansatz into the Helmholtz equation converts it into the following the eigenvalue problem
\begin{equation}
    \left[\dfrac{\partial}{\partial x^2} + \dfrac{\partial}{\partial y^2} +k^2n_0^2(x,y)\right]|\phi_j\rangle = \beta_j^2 |\phi_j\rangle
\end{equation}
with eigenvalues $\beta_j^2$; note the structural similarity of this equation and the time-independent Schrödinger equation. More general solutions can be expanded in terms of $|\phi_j\rangle$, the eigenmodes.
\\\\
In an ideal waveguide, power does not transfer between eigenmodes. However, cross-coupling can be induced by a weak perturbation in the refractive index (e.g. a bend in an optical fiber). Define the perturbed refractive index profile as 
\begin{equation}\label{eq:perteps}
    n^2(x,y,z) =  n_0^2(x,y) +\delta \epsilon(x,y,z)
\end{equation}
where $\delta\epsilon(x,y,z)$ is a weak function of $x$,$y$ in accordance with weak guidance, as well as $z$. The perturbed Helmholtz equation is 
\begin{equation}
     \left[\nabla^2  + k^2n_0^2 + k^2 \delta\epsilon\right] | \psi\rangle = 0.
\end{equation}
The CMT uses the following ansatz for the perturbed Helmholtz equation.
\begin{equation} \label{eq:ansatz1}
\begin{split} 
    |\Psi\rangle &= \sum_j a_j(z) \, e^{i\beta_j} |\phi_j\rangle.
\end{split}
\end{equation}
The above essentially claims that any propagating wavefront can be expressed as a linear combination of the ideal and $z$-invariant eigenmodes, modulated by $z$-varying coefficients, motivating the name ``fixed-basis'' CMT. Further assume that the mode coefficients $a_j(z)$ vary slowly compared to $\beta_j$ --- this approximation is covered in more detail in \S\ref{ssec:SVEA}. This yields
\begin{equation}
\begin{split}
    \left[ \dfrac{\partial^2}{\partial z^2}+ \nabla^2_\perp +k^2n_0^2 + k^2 \delta \epsilon\right]\sum_j a_j(z) \, e^{i\beta_jz} |\phi_j\rangle &= 0\\
    \sum_j \left[2i\beta_j\dfrac{d a_j}{dz}e^{i\beta_j z}|\phi_j\rangle    
    +k^2\delta \epsilon a_j e^{i\beta_j z}|\phi_j\rangle \right] &= 0
\end{split}
\end{equation}
where $\nabla_\perp \equiv \partial^2/\partial x^2 + \partial^2 / \partial y^2$.
Complete an inner product by left-multiplying with $\exp(-i\beta_i z)\langle \phi_i|$:
\begin{equation}
    \begin{split}
        2 i \beta_i \dfrac{d a_i}{dz}  + \sum_j a_j k^2 e^{i(\beta_j-\beta_i)z}\langle \phi_i |\delta \epsilon|\phi_j\rangle &= 0.
    \end{split}
\end{equation}
The coupling coefficients $\kappa_{ij}$ are defined as 
\begin{equation}
    \kappa_{ij} = \dfrac{k^2}{2 \beta_i}\langle \phi_i |\delta \epsilon|\phi_j\rangle.
\end{equation}
The definition is conventional and may differ from text to text. The mode coefficients of the propagating wavefront in the perturbed waveguide solve the following coupled system of differential equations:
\begin{equation}
    \dfrac{d a_i}{d z} = i\sum_j e^{i(\beta_j-\beta_i)z}\kappa_{ij} a_j.
\end{equation}
Appendix \ref{ap:2mode} solves the above for the simple case of a two-mode waveguide (e.g. a $2\times 2$ directional coupler).

\section{Coupled-mode theory for slowly varying waveguides}\label{ssec:deriv2}
The fixed-basis CMT fails for waveguides whose eigenbases (i.e. guided modes) vary with $z$. One such waveguide is the PL, which transitions slowly from a few-mode step-index fiber geometry to a multi-core fiber geometry. Even in the absence of perturbations, cross-coupling can be induced by the changing structure of the waveguide; in other words, the change in the waveguide structure for two nearby values of $z$ is a weak perturbation in the sense of \S\ref{ssec:deriv}. 
\\\\
To study such a waveguide, I apply the CMT to solve a differential equation of the form
\begin{equation} \label{eq:genform}
    D_z |\Psi\rangle + A(z) |\Psi\rangle = 0
\end{equation}
where $D_z$ is a linear differential operator with respect to parameter $z$ and the operator $A(z)$ varies slowly with $z$; for more details on this criterion, see \S\ref{ssec:parax}-\ref{ssec:adi}. $A(z)$ admits an ``instantaneous'' eigenbasis $|\xi_j(z)\rangle$ with eigenvalues $\lambda_j(z)$:
\begin{equation}
    A(z) |\xi_j(z)\rangle  = \lambda_j(z) |\xi_j(z)\rangle.
\end{equation}
Going forward I will indicate $z$ dependence with a superscript, i.e. $|\xi_j^z\rangle \equiv |\xi_j(z)\rangle$. As before, the above form can be obtained under weak guidance, which reduces Maxwell's equations to a set of decoupled scalar Helmholtz equations. This gives the identifications
\begin{equation}
\begin{split}
    D_z &= \dfrac{\partial^2}{\partial z^2} \\
    A(z) &= \left[\nabla_\perp^2+k^2n^2(x,y,z)\right].
\end{split}
\end{equation}
This is not the only identification possible; see \S\ref{ssec:pol}, which applies the CMT to polarized plane waves. The operator $A(z)$ has the eigenvalue equation 
\begin{equation}
    A(z)|\xi_j^z\rangle = \beta_j^2(z)\, |\xi_j^z\rangle
\end{equation}
where the $\lambda_j(z) \equiv \beta_j^2(z)$ are the eigenvalues. In the varying-basis CMT, solutions are expanded in $|\xi_j^z\rangle$. An ansatz is (c.f. equation \ref{eq:ansatz1}):
\begin{equation} \label{eq:ansatz2}
    |\Psi\rangle = \sum_j a_j(z) e^{i\int_0^z \beta_j(z')dz' }|\xi^z_j\rangle.
\end{equation}
Plugging the above back into the Helmholtz equation and completing an inner product with $\langle \Psi_i | \equiv e^{-i\int_0^z \beta_i(z')dz' } \langle \xi_i^z| $ yields:
\begin{equation}\label{eq:coupled_full}
\begin{split}
    0 &= 2i\beta_i \dfrac{d a_i}{d z} + ia_i \dfrac{d \beta_i}{d z}  \\
    &+ \sum_j e^{i\int_0^z (\beta_j-\beta_i)dz' } \left[ 2 \dfrac{d a_j}{d z} + 2ia_j\beta_j \right]\langle \xi_i^z|\dfrac{\partial}{\partial z} | \xi_j^z\rangle \\
    &+ \dfrac{d^2 a_i}{d z^2} + \sum_j a_j e^{i\int_0^z (\beta_j-\beta_i)dz' }\langle \xi_i^z|\dfrac{\partial^2}{\partial z^2} | \xi_j^z\rangle. 
\end{split}
\end{equation}
The following sections review common approximations to the above coupled-mode equations.
\subsection{Paraxial approximation}\label{ssec:parax}
The paraxial approximation is usually stated as: 
\begin{equation}
    \left| \dfrac{\partial^2 \xi_j}{\partial z^2} \right| \ll \beta_j \left| \dfrac{\partial \xi_j}{\partial z} \right|.
\end{equation}
This is equivalent to assuming that the propagation direction of the wavefront is close to the optical axis of the waveguide.\footnote{There are so-called ``wide-angle'' approximations (e.g. \citealt{Hadley:92}) which can be used to relax the assumption of paraxiality.}
This approximation removes all second-order derivatives of $|\xi\rangle$, leaving:
\begin{equation}\label{eq:coupled1}
\begin{split}
    0 &= 2i\beta_i \dfrac{d a_i}{d z} + ia_i \dfrac{d \beta_i}{d z}  \\
    &+ \sum_j e^{i\int_0^z (\beta_j-\beta_i)dz' } \left[ 2 \dfrac{d a_j}{d z} + 2ia_j\beta_j \right]\langle \xi_i^z|\dfrac{\partial}{\partial z} | \xi_j^z\rangle 
    + \dfrac{d^2 a_i}{d z^2}.
\end{split}
\end{equation}

\subsection{Slowly varying envelope}\label{ssec:SVEA}
The slowly varying envelope approximation (SVEA) claims:
\begin{equation}
    \left| \dfrac{d^2 a_j}{d z^2} \right| \ll \beta_j \left| \dfrac{d a_j}{d z} \right| \ll \beta_j^2|a_j|.
\end{equation}
In other words, the SVEA claims  that coupling between the instantaneous eigenmodes is weak, so that the $a_j(z)$ vary slowly compared to $\exp(i\beta_j z)$. I define the coupling coefficients $\Tilde{\kappa}_{ij}$ as
\begin{equation}
    \Tilde{\kappa}_{ij} \equiv \langle \xi_i^z|\dfrac{\partial}{\partial z} | \xi_j^z\rangle. 
\end{equation}
An analogous version of the coupling coefficients in the case of vector propagation are given in \citep{Jin:10}. Under the SVEA, the coupled-mode equations can be written as 
\begin{equation} \label{eq:coupled2}
    \dfrac{d a_i}{d z} = -\dfrac{1}{2} \dfrac{d \ln \beta_i}{d z}a_i - \sum_j  \dfrac{\beta_j }{\beta_i}e^{i\int_0^z (\beta_j-\beta_i)dz' }\Tilde{\kappa}_{ij}  a_j.
\end{equation}
In general, the above form is what should be used when modelling slowly varying waveguides, though the $d \ln \beta_i /d z$ term may be negligible in the case of weak guidance. When $A(z)$ is self-adjoint (or Hermitian, in our case), the eigenbasis can be chosen to be purely real, and $\Tilde{\kappa}_{ij}$ is a real antisymmetric matrix.
\\\\
For power transfer between modes $i$ and $j$ to occur, $\Tilde{\kappa}_{ij}$ must be large and $\beta_j-\beta_i$ must be small. If the latter condition is not met, the sign of the coupling will rapidly flip with $z$ and power transfer will be negligible.
\subsection{Adiabatic approximation}\label{ssec:adi}
In the adiabatic approximation, we assume that the coupling coefficients are negligible, so that the modal evolution is decoupled: 
\begin{equation}
    \dfrac{d a_i}{d z} = -\dfrac{1}{2} \dfrac{d \ln \beta_i}{d z}a_i.
\end{equation}
This approximation is analogous to the adiabatic approximation of quantum mechanics (e.g. \citealt{Taras} who applies this approximation to waveguide couplers); also, see \S\ref{ssec:geophase}. The general solution is
\begin{equation}
    a_i(z) = a_i(0) \sqrt{\dfrac{\beta_i(0)}{\beta_i(z)}}
\end{equation}
and the full solution for a propagating wavefront is 
\begin{equation}
    |\Psi(z)\rangle = \sum_j a_j(0)\sqrt{\dfrac{\beta_j(0)}{\beta_j(z)}} e^{i\int_0^z \beta_j(z') dz'}|\xi^z_j\rangle
\end{equation}
which is reminiscent of the Wentzel–Kramers–Brillouin (WKB) approximation for a particle in a slowly varying potential; see \S\ref{ssec:wkb}. While the SVEA implicitly assumes that $|\Tilde{\kappa}_{ij}| \ll \beta_i$ for all $i \neq j$, the condition for adiabaticity is even more stringent \citep{BornFock}: 
\begin{equation}\label{eq:adcond}
    |\Tilde{\kappa}_{ij}| \ll |\beta_i-\beta_j|.
\end{equation}
This condition\footnote{While not strictly sufficient, this condition holds in ``most'' cases, and is often used in quantum mechanics \citep{cond}.} can be loosely motivated by noting that the antiderivative of $\exp [i \int_0^z (\beta_j-\beta_i)dz']$, which appears in the coupled-mode equations \ref{eq:coupled2}, is of order $1/|\beta_i-\beta_j|$. In terms of the refractive index profile, the adiabatic criterion can be recast using perturbation theory (e.g. \S\ref{ssec:pert}) as 
\begin{equation}
    k^2 \Bigg| \dfrac{\langle \xi^z_i|\dfrac{\partial n^2}{\partial z}|\xi^z_j\rangle}{\beta_i^2-\beta_j^2} \Bigg|\ll |\beta_i-\beta_j| \,\, ; \,\,i \neq j.
\end{equation}
In the case of degeneracy, the adiabatic approximation breaks down: {\it there is no rate of change slow enough to prevent cross-coupling between degenerate modes.}

\subsection{Example applications of the varying-basis CMT}
To reiterate, the prior analysis suggests that light moves through a slowly-varying waveguide in co-propagating modes which cross-couple depending on mode shapes and propagation constants. The evolution of the amplitudes of these modes is controlled by a system of ordinary differential equations. This allows us to build an intuition for how light propagates through some astrophotonic devices. I provide some examples:
\paragraph{\textbf{$2\times 2$ Directional coupler}} As shown in Figure \ref{fig:intro}a, this device has two single-moded channels embedded in an optical substrate. The two channels are initially well-separated, then are slowly brought closer together, kept close over some ``coupling length'', and then slowly brought apart. When light is injected into one input, the device acts like a beamsplitter; the power ratio of the split is determined by the coupling length. One explanation of how this splitting works is given in \S\ref{ap:2mode}, using the fixed-basis CMT, but the varying-basis CMT provides a different perspective. When the two channels are very far apart, the instantaneous eigenmodes are indeed just the fundamental modes of each channel. However, when the two channels are close, the instantaneous eigenmodes look like the symmetric and antisymmetric combinations of each channel's fundamental mode. When light is launched in one input of the coupler, it is gradually split between the symmetric and antisymmetric mode. Then, as the light propagates down the coupling length, power is observed to oscillate between the two channels: this is due to the beating of the symmetric and antisymmetric mode, which have different propagation constants. The oscillation slows and eventually stops when the channels are separated again. The relevant approximations are paraxiality and the SVEA; an example device is numerically simulated in \S\ref{ssec:dicoupler}. 

\paragraph{\textbf{Mode-selective PL}}
A mode-selective PL couples light from each mode at its entrance into a separate single-mode output. In other words, a mode-selective PL has no cross-coupling between its modes, and thus should satisfy the adiabatic condition, along with paraxiality and the SVEA. One way to ensure negligible cross-coupling, mentioned in \S\ref{ssec:SVEA}, is to give each mode a sufficiently distinct propagation coefficient $\beta_i$. \citealt{mspl} achieves this by forming the PL from single-mode fibers with different core sizes; another option would be differentiate the cores in refractive index. Conversely, a non-mode-selective PL uses cores which are identical in size and refractive index. As a result, there is a region of the PL (particularly towards the larger end of the taper) where the co-propagating PL modes become degenerate and can rapidly cross-couple. In other words, the larger end of an $N$-port PL acts like an $N\times N$ directional coupler; this is also why the propagation characteristics of fabricated PLs are difficult to control. A non-mode-selective PL is simulated in \S\ref{ssec:PL}.

\paragraph{\textbf{Fiber Bragg gratings}}
Briefly, I mention that the CMT can be extended to structures such as fiber Bragg gratings, which use a rapid variation within the refractive index structure of an optical fiber to reflect narrow spectral bands. To apply the CMT, we must include both the forward and backward propagating waves when expanding the electric field within the waveguide; see \citealt{McCall:00}. Note that despite the rapid variation, a fiber Bragg grating may still be thought of as slowly varying in the sense of the SVEA due to the low index contrast of the variation.

\section{Connections to quantum mechanics}\label{sec:qm}
Several studies have illustrated how certain optical waveguides have quantum mechanical counterparts: for instance, the PL and the Kronig-Penney model for electrons in a periodic potential well \citep{Leon-Saval2,joss,Kittel}. From another perspective, the fixed-basis CMT is essentially an application of time-dependent perturbation theory \citep{CH}, with the time $t$ replaced with longitudinal coordinate $z$. In this section, I demonstrate that the varying-basis CMT is also connected to the quantum mechanics. In \S\ref{ssec:wkb}, I show that the first-order WKB approximation and the CMT under the adiabatic approximation give the same solution when applied to the Helmholtz equation. In \S\ref{ssec:pert}, I review how time-independent perturbation theory can be used to compute the coupling coefficients $\Tilde{\kappa}_{ij}$ \citep{Shi:16}. Lastly, in \S\ref{ssec:geophase}, I show how geometric phase, which appears in quantum mechanics under the adiabatic approximation, is a natural result of the CMT.

\subsection{WKB approximation}\label{ssec:wkb}
The WKB approximation is closely related to the varying-basis CMT, and in fact motivates the ansatz in equation \ref{eq:ansatz2}. To see this, I present a brief overview of the WKB approximation. Suppose we have a differential equation 
\begin{equation}
    \dfrac{dy}{dx} = f(x,y).
\end{equation}
The above can be recast in integral form as 
\begin{equation}
\begin{split}
    y(x) &= \int_{x_0}^{x} f(x',y(x')) dx'+C \\
    C &= y(x_0).
\end{split}
\end{equation}
To solve this integral equation we can provide an initial guess for $y(x)$, denoted $y_0(x)$. Substituting this guess back into the integral equation gives a more accurate guess $y_1(x)$. This process can be repeated:
\begin{equation}
    y_n(x) = \int_{x_0}^{x} f(x',y_{n-1}(x')) dx'+ C.
\end{equation}
In both quantum mechanics and optics (under the scalar approximation), we deal the Helmholtz equation:
\begin{equation}\label{eq:genwave}
    \dfrac{d^2 \psi}{d x^2} + k^2(x)\psi = 0. 
\end{equation}
For $k(x)=k$ constant, solutions are the plane waves: $\psi(x) = \exp(\pm ikx)$. If $k(x)$ is allowed to vary, then we might replace $kx$ with $\int k(x) dx$. This is gives us the 0th order approximation
\begin{equation}
    \psi_0(x) = e^{\pm i\int k(x) dx}.
\end{equation}
which is valid for $\left|d k/dx\right| \ll k^2$. I now apply the method of successive approximations to solve the wave equation \ref{eq:genwave}. Assume a solution of the form $\psi(x) = \exp\left[i S(x)\right]$.
Plugging this into equation \ref{eq:genwave} yields
\begin{equation} \label{eq:wkbdiffeq}
    i S'' - (S')^2 + k^2 = 0.
\end{equation}
Manipulate the above to derive the series relation for $S_n$, and insert the 0th order approximation. This yields the first-order solution
\begin{equation}
    \psi_1(x) \propto \dfrac{1}{\sqrt{k(x)}}e^{\pm i\int k(x)\, dx}.
\end{equation}
For more details, see appendix \ref{ap:wkb}. This form is the same as what was derived in \S\ref{ssec:deriv} under the adiabatic approximation, and justifies our ansatz \ref{eq:ansatz2}.

\subsection{Time-independent perturbation theory}\label{ssec:pert}
Besides using finite differences, the varying-basis coupling coefficients $\Tilde{\kappa}_{ij}$ can also be estimated from the fixed-basis coefficients $\kappa_{ij}$ using time-independent perturbation theory --- or in our case, $z$-independent perturbation theory. Define the refractive index structure of our waveguide as $n(x,y,z)$. We can treat the difference in $n^2(x,y,z)$ between two values of $z$, separated by some small $\delta z$, as a perturbation $\delta \epsilon$:  
\begin{equation}
    \delta \epsilon = n^2(x,y,z+\delta z) - n^2(x,y,z).
\end{equation}
Following perturbation theory, we treat $k^2\delta \epsilon$ as a perturbation to the operator $A \equiv \nabla_\perp^2 + k^2 n^2(x,y,z)$ in the eigenvalue equation $A|\xi_j^z\rangle = \beta_j^2(z)|\xi_j^z\rangle$. The first-order approximation of the perturbed eigenmode is \citep{Sakurai}
\begin{equation}
    |\xi_j^{z+\delta z}\rangle \approx |\xi_j^{z}\rangle + \sum_{i\neq j} \dfrac{\langle \xi_i^z|k^2\delta \epsilon|\xi_j^z\rangle}{\beta^2_j-\beta^2_i} |\xi_i^z \rangle.
\end{equation}
Therefore, the coupling coefficient matrix can be approximated as 
\begin{equation}
\begin{split}
    \langle \xi^z_k |\dfrac{\partial}{\partial z}|\xi^z_j\rangle &\approx \dfrac{1}{\delta z} \sum_{i\neq j} \dfrac{\langle \xi_i^z|k^2\delta \epsilon|\xi_j^z\rangle}{\beta^2_j-\beta^2_i}\langle \xi_k^z |\xi_i^z \rangle\\
&=
\begin{cases}
    \dfrac{\langle \xi_k^z|k^2\delta \epsilon/\delta z|\xi_j^z\rangle}{\beta^2_j-\beta^2_k}, & k \neq j\\
    0, & k = j.
\end{cases}
\end{split}
\end{equation}
In the limit $\delta z \rightarrow 0$, we have $\delta \epsilon/\delta z \rightarrow \partial n^2 /\partial z $. This formula links the coupling matrix of the varying-basis CMT in \S\ref{ssec:deriv2} to the that of the fixed-basis CMT in \S\ref{ssec:deriv}.
Degeneracy can be treated if it is lifted by the perturbation, using degenerate perturbation theory. However, a degeneracy enforced by a symmetry in the waveguide design may not be broken by the perturbation. The resulting coupling between degenerate modes can be seen as a result of a non-abelian geometric phase, and is discussed further in \S\ref{ssec:WZ}.

\subsection{Adiabatic approximation: geometric phase}\label{ssec:geophase}
In this section I show that the varying-basis CMT naturally includes geometric phase, which was first observed in studies of polarized light by Pancharatnam \citep{Pan} and has been widely studied in the context of quantum mechanics under the adiabatic approximation (e.g. the Aharanov-Bohm effect \citealt{AB,Berry}). Several well-known examples of geometric phase in optics include the spin redirection phase, in which the polarization orientation is rotated by out-of-plane propagation \citep{spin}; the Gouy phase, which occurs for a Gaussian beam propagating through focus \citep{Gouy}; and the Pancharatnam-Berry (PB) phase, which can occur in birefringent materials. Taking the PB phase as an example, I first show how geometric phase appears in the CMT, and then discuss how non-adiabatic evolution produces the non-abelian Wilczek-Zee phase.
\subsubsection{Pancharatnam-Berry phase}\label{ssec:pol}
Following \S\ref{ssec:deriv2}, the coupled-mode equations are 
\begin{equation}\label{eq:cmd2}
    \dfrac{d a_i}{d z} = -\dfrac{1}{2}\dfrac{d \ln \beta_i}{d z} a_i  - \sum_j \dfrac{\beta_j}{\beta_i} \langle \zeta_i^z|\dfrac{\partial}{\partial z} | \zeta_j^z\rangle  a_j
\end{equation}
where the eigenbasis has been temporarily modified as 
\begin{equation}
    |\zeta^z_j \rangle \equiv e^{i\int_0^z \beta_j \,dz' }|\xi^z_j\rangle.
\end{equation}
Using the above form and splitting the summation gives 
\begin{equation}\label{eq:coupledgeom}
    \dfrac{d a_i}{d z} = -\dfrac{1}{2}\dfrac{d \ln \beta_i}{d z} a_i - \sum_{j\neq i} \dfrac{\beta_j}{\beta_i} \langle \zeta_i^z|\dfrac{\partial}{\partial z} | \zeta_j^z\rangle  a_j -  \langle \xi_i^z|\dfrac{\partial}{\partial z} | \xi_i^z\rangle  a_i.
\end{equation}
Under the adiabatic approximation, all off-diagonal terms are set to 0, and the coupled equations can be solved via the matrix exponential:
\begin{equation}
\begin{split}
    a_i(z) &= \exp \left[ - \int_0^z \left( \dfrac{1}{2\beta_i} \dfrac{d \beta_i}{d z'} + \langle \xi_i^{z'} |\dfrac{\partial}{\partial z'} | \xi_i^{z'} \rangle  \right) dz' \right] \\ 
    &\propto \dfrac{1}{\sqrt{\beta_i(z)}} \exp \left[ - \int_0^z   \langle \xi_i^{z'} |\dfrac{\partial}{\partial z'} | \xi_i^{z'} \rangle   dz' \right].
\end{split}
\end{equation}
The matrix exponential is applicable here because the adiabatic terms in equation \ref{eq:coupledgeom} can be represented as diagonal matrices, which commute. Plugging $a(z)$ back into the original ansatz \ref{eq:ansatz2}, the overall wavefront evolves as
\begin{equation}
    |\Psi_j(z)\rangle \propto \dfrac{1}{\sqrt{\beta_j(z)}} e^{i \phi_j(z)} e^{i\theta_j(z)}|\xi^z_j\rangle
\end{equation}
where $\phi_j(z)$ and $\theta_j(z)$ are defined as 
\begin{equation}
\begin{split}\label{eq:geoph}
    \phi_j(z) &\equiv i\int_0^z   \langle \xi_i^{z'} |\dfrac{\partial}{\partial z'} | \xi_i^{z'}\rangle dz'\\
    \theta_j(z) &\equiv \int_0^z \beta_j(z')dz'. 
\end{split}
\end{equation}
It can be shown that $\phi_j$ is real.\footnote{ The operator $\partial/\partial z$ is anti-Hermitian, so the diagonal of the operator in matrix representation (the bracketed term in equation \ref{eq:geoph}) is purely imaginary. This can also be proven using the product rule.} The term $\theta_j(z)$ is the dynamical phase, while $\phi_j(z)$ is identified as the PB phase when the system evolves in a loop, i.e. when $|\xi_j^0 \rangle = |\xi_j^z\rangle$. In quantum mechanics, $\phi_j(z)$ is often expressed as a contour integral in some parameter space with parameters $\lambda_\mu$. Then, $|\xi_i(z)\rangle =|\xi_i(\lambda_\mu(z)\rangle$, and the PB phase is 
\begin{equation}\label{eq:geomphase}
    \phi_j(z) = i\oint_\gamma \sum_\mu  \langle \xi_j |\dfrac{\partial}{\partial \lambda_\mu}| \xi_j \rangle  d\lambda_\mu \equiv \oint_\gamma \bm{\mathcal{A}}\cdot d\bm{\lambda}
\end{equation}
where $\gamma(z) = (\lambda_1(z),\lambda_2(z),...)$ denotes a closed path in $\lambda_\mu$ parameter space; for the last definition, I have switched to vector notation. The vector $\bm{\mathcal{A}}$ is the Berry or Aharanov-Anandan connection. The curl of the connection $\bm{\mathcal{A}}$ is the Berry curvature, denoted $\bm{\Omega}$.\footnote{The connection and curvature can be alternatively written using differential forms. Specifically, we may write the connection as the one-form $\mathcal{A}=\langle \xi_j |d| \xi_j\rangle$, where $d$ is the exterior derivative. The connection is the two-form $\Omega=d\mathcal{A}$. }
\\\\
When the eigenmodes $|\xi^z_j\rangle$ are real-valued, as is the case under the scalar approximation (neglecting polarization), the Berry connection is 0 and no PB phase can be accrued.\footnote{For a mode $\xi$, if we have $\xi=\Bar{\xi}$ then $\int \Bar{\xi} \dfrac{\partial}{\partial z} \xi \,dx dy = \int \xi \dfrac{\partial}{\partial z} \xi \,dx dy = \frac{1}{2}\dfrac{\partial}{\partial z} \int \xi^2\, dx dy = 0. $} Thus, to see PB phase, the eigenvectors must be complex --- equivalently, the operator $A(z)$ cannot be self-adjoint. This is the case for the propagation of polarized plane waves through a slowly varying birefringent medium. For instance, represent a polarized plane wave $|P\rangle$ in terms of its normalized Jones vector $|p\rangle$:
\begin{equation}
\begin{split}
        |P\rangle &= e^{i(\beta z-\omega t)} e^{i\gamma} |p\rangle \\
        |p\rangle &\equiv     \begin{bmatrix}
    \cos\chi/2 \\
    e^{-i\psi}\sin\chi/2  
    \end{bmatrix}. 
\end{split}
\end{equation}
Here, $\gamma,\psi \in [0,2\pi]$, $\chi \in [0,\pi]$, and $\beta$ is the propagation constant. The parametrization for $|p\rangle$ is also used for rank-1 spinors in quantum mechanical two-state systems, and applies to anything that can be represented as a unit vector in $\mathbb{C}^2$. The angle $\gamma$ represents absolute phase; for visualization, the ``azimuthal'' angle $\psi$ and ``polar'' angle $\chi$ are often mapped onto the surface of the so-called Poincar\'e sphere\footnote{Or the Bloch sphere, in the study of two-state quantum mechanical systems.}, which represents all possible states of the polarization ellipse over its surface (\citealt{sphere} provides a nice explanation).
\\\\
Following studies of quantum-mechanical two-state systems, the propagation of a polarized plane wave $|P\rangle$ through a homogeneous birefringent material that slowly varies with $z$ can be represented by the differential equation 
\begin{equation}
\begin{split}
    \dfrac{\partial}{\partial z}|p\rangle &= \mathfrak{J}(z) |p\rangle \\
\end{split}
\end{equation}
where $\mathfrak{J}(z)$ is a $z$-dependent anti-Hermitian matrix, i.e. a linear combination of the matrices $i\sigma_j$, where the $\sigma_j$ are the Hermitian Pauli spin matrices. We can motivate the above by noting that the Jones matrices, which represent the optical transfer matrices for homogeneous slabs of birefringent material, are a representation of the special unitary group SU(2), whose Lie algebra  is $\mathfrak{su}(2)$, which in turn has the basis $i\sigma_j$. The evolution of $|p\rangle$ is parameterized by the path $\gamma(z) = [\psi(z),\chi(z)]$. 
\\\\
Comparing the above equation with the form assumed in \S\ref{ssec:deriv2}, we identify $D(z) = \partial/\partial z$ and $A(z)=-\mathfrak{J}(z)$; the differential equation governing the evolution of the wavefront is first-order, not second-order. This simplifies the application of the CMT, removing the ``WKB'' term from the coupled-mode equations \ref{eq:cmd2}; our prior results for geometric phase still hold. The Berry connection for polarization states, $\bm{\mathcal{A}}_{\rm pol}$, has components
\begin{equation}
\begin{split}
    \mathcal{A}_{\rm pol,\chi} &= 0 \\
    \mathcal{A}_{\rm pol,\psi} &= \dfrac{\sin^2 \chi/2}{\sin \chi}.
\end{split}
\end{equation}
The PB phase can be computed using Stokes' theorem:
\begin{equation}
    \phi = \oint_\gamma \bm{\mathcal{A}}\cdot d\bm{\lambda} = \iint_\mathcal{S}  \left(\nabla \times \bm{\mathcal{A}} \right)\cdot d\bm{\mathcal{S}} = \dfrac{1}{2}\iint_\mathcal{S} \, d\mathcal{S}.
\end{equation}
Here, $\mathcal{S}$ denotes the surface of the unit 2-sphere (the Poincar\'e sphere) bound by the closed curve $\gamma$, and $d\mathcal{S}$ is an area element of the sphere; the curvature is $\Omega = 1/2$.\footnote{This can be derived using the definition of the connection from equation \ref{eq:geomphase} and the formula for the curl in spherical coordinates.} Thus, we recover that the PB phase accumulated by a polarized plane wave is half the solid angle enclosed by the path it takes over the Poincar\'e sphere \citep{polgeom}.
\\\\
The PB phase can be more generally understood as a holonomy in the context of fiber bundles \citep{fib}. The full polarization state $\exp[-i \psi] |p\rangle$ ``lives'' in a unit 3-sphere, denoted $S^3$, which can be represented as a fiber bundle whose base space is the unit 2-sphere $S^2$ and whose fiber space is the unit circle $S^1$ through the Hopf fibration\footnote{Equivalently, the fiber space is the unitary group $U(1)$.} \citep{Hopf}. The base space --- the Poincar\'e sphere --- records the orientation and eccentricity of the polarization ellipse, while the fibers record absolute phase. Closed paths on the base space may be lifted by the fiber bundle connection --- the Berry connection --- to open paths in the total space of polarization states $S^3$. These open paths start and end on the same fiber, and the fiber-wise ``distance'' between the start and end points of such paths encodes the geometric phase accumulated by cyclic evolution about the Poincar\'e sphere.

\subsubsection{Degenerate adiabatic approximation: Wilczek-Zee phase}\label{ssec:WZ}
Though wavefronts in the scalar approximation accumulate no PB phase, they can accumulate Wilczek-Zee (WZ) phase so long as the eigenbasis is at least partially degenerate. The WZ phase generalizes the PB phase, which was derived under the adiabatic approximation, to degenerate eigenspaces \citep{WZ}. To see how the WZ phase appears in the CMT, consider solving a simple form of the coupled-mode equations \ref{eq:cmd2}, assuming degeneracy of all eigenmodes, dropping the WKB term, and discretizing the interval $[0,z]$ at $N$ evenly-spaced values $z_n = n\, \Delta z$, with $n=0,1,2...N$ and $z_N = z$.  Denote the series of matrices $M_{ij}^{z_n}$ as 
\begin{equation}
    M_{ij}^{z_n} \equiv \langle \zeta_i^z|\dfrac{\partial}{\partial z} | \zeta_j^z\rangle \Big|_{z=z_n}.
\end{equation}
The $M^{z_n}$ form a non-commuting group. The solution to the coupled-mode equations can be cast as the limit:
\begin{equation}
\begin{split}
    \vec{a}(z) &= \lim_{\Delta z \rightarrow 0} \left(e^{M^{z_N}\Delta z} \,... \, e^{M^{z_1}\Delta z} e^{M^{z_0}\Delta z}\right)\vec{a}(0) \\
    &\equiv  \mathcal{P} \left\{ e^{ \int_0^z M(z') dz'} \right\} \vec{a}(0)
\end{split}
\end{equation}
where I have used the notation $\mathcal{P} \{ \exp { \int_0^z M(z') dz'}\}$ to denote the path-ordered exponential of $M(z)$; in general, we cannot take the usual matrix exponential because the $M^z_{ij}$ do not commute, and so we must take our matrix exponential in short steps of $\Delta z$ in a particular order --- hence why the WZ phase is termed non-abelian. The WZ phase arises from the path-ordered exponential of $M_{ij}(z)$, and controls the variation in the phases and relative weights of the degenerate eigenmodes. In the language of fiber bundles, the WZ phase can be seen as the holonomy of a vector bundle, while abelian geometric phases are specifically holonomies of line bundles \citep{ZHANG20231}.
\\\\
Geometric phases both abelian and non-abelian are of particular interest in quantum computing due their anticipated resilience against noise \citep{ZHANG20231}, and have been experimentally realized by manipulating qubit systems \citep{nonabel}. While not completely analogous, there is a similar need for noise-robust and achromatic optical interferometry, especially for the purpose of astronomical nulling. Achromatic nullers using rotational-shearing interferometers have been implemented \citep{Serabyn:99,Tavrov:05}; it would be interesting to see if a photonic equivalent could also be made by applying geometric phase manipulations within a waveguide. As a monolithic waveguide, such a device would be more stable and compact.

\section{Numerical implementation: \texttt{cbeam}}\label{sec:prax}
\begin{figure}
    \centering
    \includegraphics[width=0.4\columnwidth]{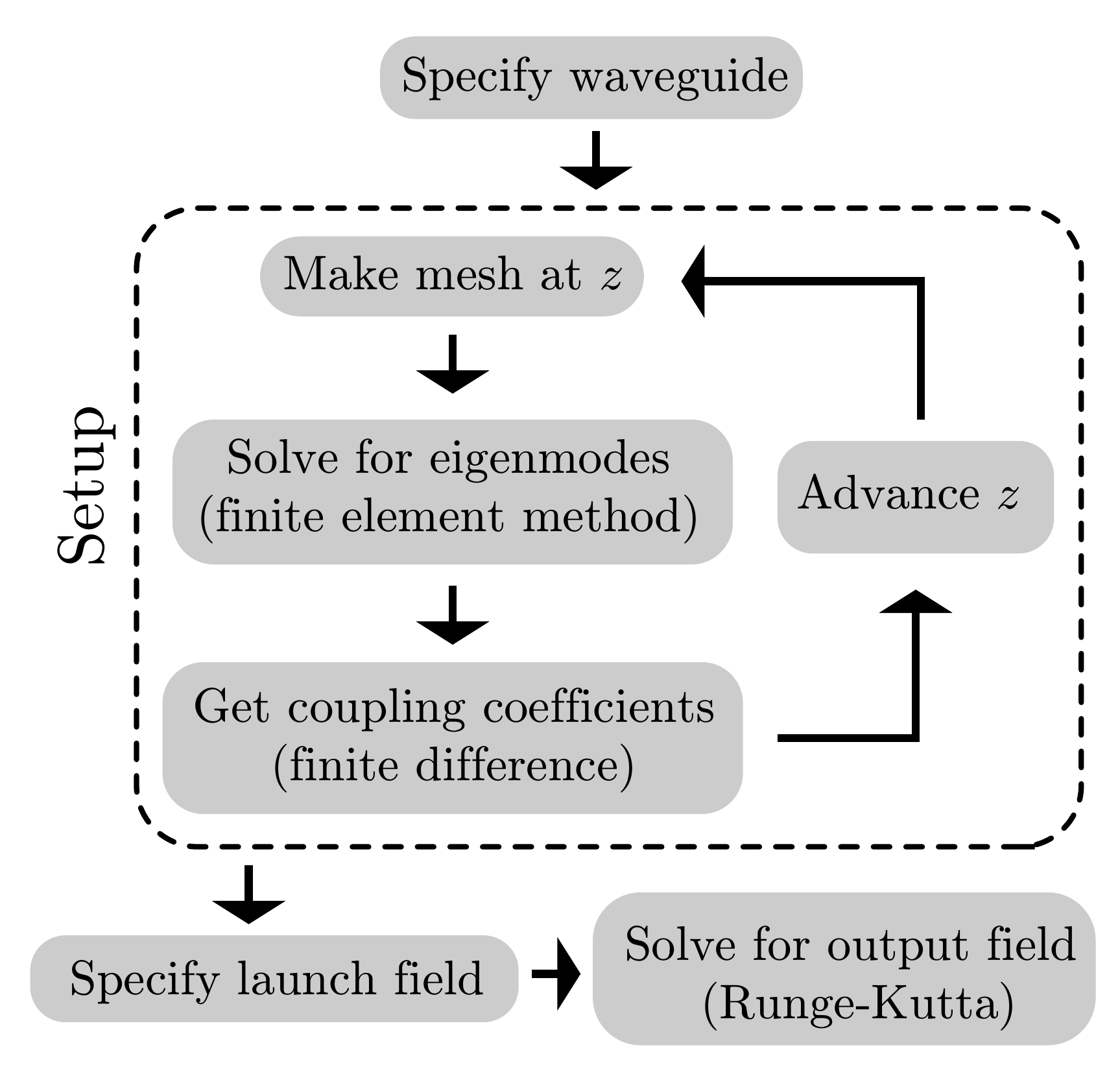}
    \caption{Block diagram of the \texttt{cbeam} package. The loop within the dashed lines takes the bulk of the simulation time but only needs to be run once per waveguide. }
    \label{fig:block}
\end{figure}
In this section I present a lightweight numerical implementation of the varying-basis CMT, written in Python and Julia, called \texttt{cbeam} \citep{cbeam}. Compared to alternate packages which use the finite-difference beam propagation method (FDBPM), \texttt{cbeam} can be more than an order of magnitude faster; see \S\ref{ssec:PL}. Figure \ref{fig:block} gives a graphical overview of \texttt{cbeam}'s structure. A user first specifies a waveguide; \texttt{cbeam} then adaptively computes the following as a function of $z$:
\begin{enumerate}
    \item the instantaneous propagation coefficients $\beta_j(z)$
    \item the instantaneous eigenmodes $|\xi_j^z\rangle$
    \item the coupling coefficients $\Tilde{\kappa}_{ij}$
\end{enumerate}
This setup step is run once per waveguide. Items 1 and 2 are computed using the finite-element method, though care must be taken in the case of degeneracy --- see \S\ref{ssec:degen}. Item 3 is computed by interpolating through the modes using cubic splines and evaluating the derivative; though a correction must be made --- see \S\ref{ssec:dif}. With these items and a launch field in hand, propagation through the waveguide becomes an initial value problem, which \texttt{cbeam} solves using an adaptive Runge-Kutta method. In \S\ref{ssec:dicoupler} and \S\ref{ssec:PL}, I use \texttt{cbeam} to simulate a directional coupler and 6-port photonic lantern. For more detailed examples, refer to the Github repository for \texttt{cbeam} and online documentation, located \href{https://github.com/jw-lin/cbeam/}{here}.

\subsection{Eigenmode continuity}\label{ssec:degen}
When eigenvalues are degenerate, the eigenbasis is not well-defined. Under such conditions, the instantaneous eigenmodes returned by a finite-element solver may not evolve continuously with $z$. To enforce continuity, \texttt{cbeam} models the $z$-dependent variation in waveguide geometry by applying a continuous spatial transformation to the nodes of the finite element mesh. This transformation is not specific to a single type of waveguide, and works by expanding or contracting regions of the finite element mesh so that triangles remain approximately similar. For example, this transformation can model a directional coupler by shrinking and expanding the space between a set of single-mode channels. Additionally, if the waveguide admits a degenerate subspace of eigenmodes throughout its entire length, \texttt{cbeam} uses least-squares to fix a degenerate eigenbasis and prevent it from rotating. The procedure is as follows: when going from $z\rightarrow z+\delta z$, apply a unitary transformation to the degenerate modes at $z+\delta z$ to ``match'' the modes at $z$. The transformation can be found via least-squares using the singular value decomposition (SVD). Define the degenerate eigenbases at $z$ and $z+\delta z$ as $|\xi_k^z\rangle $ and $|\xi_k^{z+\delta z}\rangle $, respectively. Next, define the overlap matrix $C$ as 
\begin{equation}
    C_{nk} \equiv  \langle \xi_n^{z+\delta z}|\xi_k^z\rangle.
\end{equation}
Construct the unitary matrix $Q$, which rotates the eigenbasis, from the SVD components of $C$:
\begin{equation}
\begin{split}
    C &= U S V^T \\
    Q &\equiv VU^T.
\end{split}
\end{equation}
Denote $M$ a matrix whose columns are the eigenmodes computed at $z+\delta z$, i.e. $|\xi_j^{z+\delta z}\rangle$. The new basis should be chosen as the columns of $M'$ with
\begin{equation}
    M' = M Q.
\end{equation}
Another complication is that eigenvalues of a general operator $A(z)$ may cross as a function of $z$, requiring a method to track eigenvalues through crossings.\footnote{There are theorems called ``avoided crossing'' or ``anticrossing'' theorems from quantum mechanics which establish when crossing may occur \citep{Landau1976Mechanics}.} In tandem with the above degeneracy correction, \texttt{cbeam} also attempts to re-sort modes so that they can be tracked through crossings. 

\subsection{Estimate of the coupling matrix}\label{ssec:dif}
To compute the coupling matrix $\kappa_{ij}$, \texttt{cbeam} interpolates the eigenmode profiles through $z$ and evaluates the derivative of the interpolation directly. However, because the eigenmodes are only defined on the mesh nodes, and because the mesh nodes move with the transformation, the derivative of the interpolation function does not precisely give the partial derivative with respect to $z$. Denoting the position of a mesh node as $\bm{p}(z) = (x(z),y(z))$, the $z$ partial derivative of a mode $|\xi_j\rangle$ at that point is
\begin{equation}
    \dfrac{\partial}{\partial z}|\xi_j\rangle = \dfrac{d}{dz} |\xi_j\rangle - \nabla_\perp |\xi_j\rangle\Big|_{\bm{p}} \cdot \dfrac{d \bm{p}}{dz}
\end{equation}
where $\frac{d}{dz} |\xi_j\rangle$ is the derivative of $|\xi_j\rangle$ that ``comoves'' with the mesh point $\bm{p}$; this is what is measured when taking the derivative of the mode's interpolation function. To compute the transverse gradient of $|\xi_j\rangle$, \texttt{cbeam} uses the ``average gradient on star method" \citep{MANCINELLI}.
\begin{figure}[ht]
    \centering
    \includegraphics[width=\textwidth]{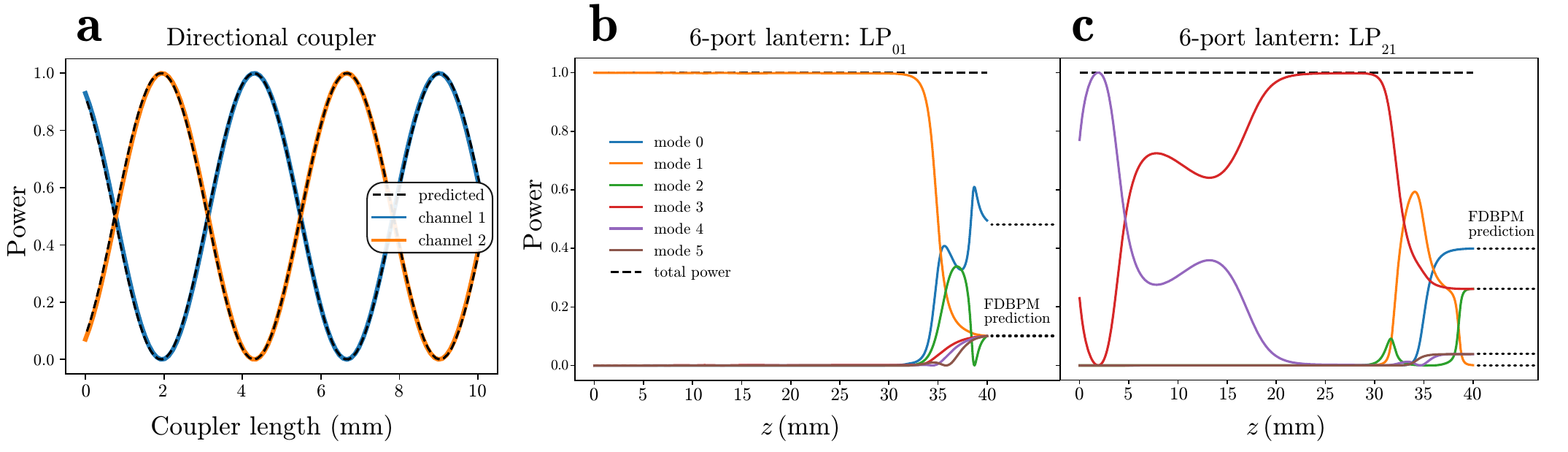}
    \caption{\textbf{a.} Output powers of a $2\times 2$ directional coupler as a function of coupling length. The dashed curves show the prediction from theory. \textbf{b, c.} The powers of the instantaneous eigenmodes in a standard 6-port PL as a function of $z$, when light is launched in the LP$_{\rm 01}$ mode (\textbf{b}) and LP$_{\rm 21}$ mode (\textbf{c}). The dotted lines past $z=40$ mm denotes the predicted lantern output from \texttt{lightbeam}, a waveguide simulation package which uses the finite difference beam propagation method.}
    \label{fig:dicoupler}
\end{figure}
\subsection{Example: directional coupler}\label{ssec:dicoupler}
Figure \ref{fig:dicoupler} (left) shows the simulated outputs of a $2\times 2$ directional coupler constructed from two single-moded circular channels which bend toward and away from each other along arctangent paths. The output power in each channel is plotted as a function of the length of the coupling region. These powers oscillate like $\sin^2$ and $\cos^2$, as expected, while the oscillation period matches a common empirical approximation \citep{TT86}, shown by the dashed curves. I include this approximation in \S\ref{ap:2mode} because some reference texts \citep{AgrawalDC,khare} contain a typographical error. 

\subsection{Example: photonic lantern}\label{ssec:PL}
The right panels of Figure \ref{fig:dicoupler} show the powers of the instantaneous eigenmodes as a function of $z$, when light is launched in the LP$_{01}$ and LP$_{21}$ modes. The modelled PL has a 5-fold rotational symmetry, with one centrally located core and five outer cores located at the points of a regular pentagon, and thus we expect that light launched in a rotationally symmetric mode should maintain a symmetric distribution at the lantern output. \texttt{cbeam} recovers this behavior: launching light into LP$_{\rm 01}$ yields equal power in all outer cores at the lantern output. Symmetry is not maintained throughout the entirety of the lantern because of the freedom of eigenbasis choice in presence of degeneracy. 
\\\\
The right panels of Figure \ref{fig:dicoupler} also show the predicted lantern port powers from \texttt{lightbeam}, a Python-based numerical package for waveguide modelling analogous to \texttt{cbeam}, which uses the FDBPM instead of the CMT. When propagating the LP$_{11}$ modes (not shown) and LP$_{21}$ modes, \texttt{cbeam} and \texttt{lightbeam} produce output power vectors that are consistent to within 0.2\% total error. For the radially symmetric modes  LP$_{01}$ and LP$_{02}$, the error is $\sim 1.5$\%. While far from a rigorous explanation, I provide one possible reason for this discrepancy. The CMT predicts small ($\lesssim 1$\%) amount of oscillatory cross-coupling at the lantern entrance between the LP$_{01}$ and LP$_{02}$ modes, which notably have the largest difference in effective index in the 6-mode group supported by the PL. This oscillatory coupling may be more difficult to model with FDBPM methods, which assign a single reference index for all propagating waves, and which accumulate more error when the chosen index chosen deviates from the effective indices of the eigenmodes \citep{bpm}. 
\\\\
Finally, I note runtime comparisons: for the fiducial 6-port PL, a single propagation in \texttt{lightbeam} takes around an hour of single-threaded computation time, and determination of the waveguide transfer matrix takes 6 propagations. With \texttt{cbeam}, the entire waveguide is modelled in around 1 minute, and subsequent propagations of any guided wavefront through the lantern take $\sim 1$s. Therefore, in this example \texttt{cbeam} is almost $400\times$ faster.

\section{Conclusion}
In this paper, I have provided an overview of coupled-mode theory, reviewing both the fixed-basis approach often used in telecommunications and the extension of the theory to waveguides which do not admit a fixed eigenbasis. I then showed how the coupled-mode formalism connects to many well-known mathematical tools from quantum mechanics, including the WKB approximation, perturubation theory, and the adiabatic approximation, which gives rise to geometric phase. Finally, I presented a numerical implementation of the varying-basis coupled-mode theory, \texttt{cbeam}, which allows for the efficient modelling of slowly varying waveguides including directional couplers and PLs. Using \texttt{cbeam}, I recover the expected behavior for fiducial examples of a $2\times 2$ directional coupler and 6-port PL.

\appendix

\section{Two-mode coupling}\label{ap:2mode}
Consider a waveguide such as a directional coupler supporting only two modes whose eigenvalues are $\beta_1^2$ and $\beta_2^2$. The two coupled-mode equations governing waveguide propagation are
\begin{equation}
    \dfrac{d}{d z} \begin{bmatrix}
        a_1 \\
        a_2
    \end{bmatrix}
    =
    i\begin{bmatrix}
        0 & \kappa_{12} e^{i (\beta_2-\beta_1)z }\\
        \kappa_{21}e^{i (\beta_1-\beta_2)z } & 0
    \end{bmatrix}
\begin{bmatrix}
        a_1 \\
        a_2
    \end{bmatrix}.
\end{equation}
To solve the above, make the following transformation:
\begin{equation}
\begin{split}
    a_1(z) &= b_1(z) e^{-i\delta z} \\
    a_2(z) &= b_2(z) e^{+i\delta z}.
\end{split}
\end{equation}
Here, $\delta\equiv (\beta_1-\beta_2)/2$. Under this transformation, the coupled system becomes:
\begin{equation}
\begin{split}
\dfrac{d}{d z}
    \begin{bmatrix}
        b_1 \\
        b_2 
    \end{bmatrix}
    &= 
    \begin{bmatrix}
        i \delta & i \kappa_{12} \\
        i \kappa_{21} & - i\delta
    \end{bmatrix}
    \begin{bmatrix}
        b_1 \\
        b_2
    \end{bmatrix}\\
    &\equiv B    \begin{bmatrix}
        b_1 \\
        b_2
    \end{bmatrix}.
\end{split}
\end{equation}
Define $\kappa_e = \sqrt{\delta^2 + \kappa_{12}\kappa_{21}}$. The solution to the coupled system is
\begin{equation}
    {\bf b}(z) = e^{zB} {\bf b}_0
\end{equation}
for initial condition ${\bf b}_0$.
Assuming ${\bf b}_0 = [1,0]$ (light is launched only in mode 1), the power in mode 2 varies as
\begin{equation}
\begin{split}
    b_2(z) &= \dfrac{\kappa_{21}e^{i\kappa_e z}- \kappa_{21}e^{-i\kappa_e z}}{2\kappa_e} \\
    &= \dfrac{i\kappa_{21}}{\kappa_e} \sin \left(\kappa_e z\right).
\end{split}
\end{equation}
Notice that up to 100\% of the power may be transferred to mode 2 if $\delta=0$: this is resonant coupling. This limit goes to 0 as the effective index difference increases. The maximum power transfer first occurs at $z=L_c=\pi/2\kappa_e$, defining the so-called coupling length $L_c$. The above shows that degenerate modes are the most susceptible to an exchange of power.
\\\\
For a symmetric directional coupler composed of two circular single-mode cores separated by a distance $d$, we may take as our idealized eigenmodes the fundamental modes supported by each core {\it when ignoring the other core}. The refractive index of the other core is treated as a perturbation $\delta \epsilon$ in the sense of equation \ref{eq:perteps}. The coupling coefficient has the following approximation \citep{TT86}:
\begin{equation}
\begin{split}
    \kappa &\equiv \kappa_{21} = \kappa_{12} \\
    &\approx \dfrac{\pi V}{2 k n_{\rm clad} a^2} \exp \left[ -2.3026\left(a + b \Tilde{d} + c \Tilde{d}^2 \right)\right]
\end{split}
\end{equation}
where $V$ is the normalized frequency parameter
\begin{equation}
    V = k a \sqrt{n_{\rm core}^2 - n_{\rm clad}^2}.
\end{equation}
Here $k$ is the free-space wavenumber; $n_{\rm core}$ and $n_{\rm clad}$ are the core and cladding indices of the coupler; $a$ is the radius of the single-mode cores; and $\Tilde{d}\equiv d/a$ is the normalized center-to-center spacing. The coefficients $a$, $b$, and $c$ are:
\begin{equation}
    \begin{split}
        a &= 2.2926 -1.5910 V +0.1668 V^2 \\
        b &= -0.3374 + 0.5321V - 0.0066 V^2 \\
        c &= -0.0076 - 0.0028 + 0.0004 V^2.
    \end{split}
\end{equation}
The sign of the last term in $c$ is sometimes erroneously flipped in the literature. This approximation is accurate within 1\% for $1.5 \leq V \leq 2.5$ and $2.0\leq \Tilde{d} \leq 4.5$.

\section{The WKB method from successive approximations}\label{ap:wkb}
To derive the first-order WKB solution, begin with the differential equation \ref{eq:wkbdiffeq}. I manipulate this equation to get the first-order derivative on the left-hand side, and then apply successive approximations as follows:
\begin{equation}
    \begin{split}
        S'(x) &= \pm\sqrt{iS''(x) + k^2(x)} \\
        S_n(x) &= \pm\int\sqrt{iS_{n-1}'' + k^2 }dx \\
        S_0(x) &= \pm\int k(x)\, dx \\
        S_1(x) &= \pm \int \sqrt{\pm ik' + k^2 } dx \\
        &\approx \pm \int k \left[1 \pm \frac{i}{2}\frac{k'}{k^2} \right] dx \\
        &= \pm \int k \, dx + \frac{i}{2}\ln(k) + C 
    \end{split}
\end{equation}
where going from line 4 to line 5 I made use of the slowly varying condition $\left|d k/dx\right| \ll k^2$ to apply a Taylor expansion. Therefore, the first-order solution is 
\begin{equation}
    \psi_1(x) = e^{iS_1(x)} \propto \dfrac{1}{\sqrt{k(x)}}e^{\pm i\int k(x)\, dx}.
\end{equation}

\section{Notes on issues with publication}
This paper started out as my personal notes on coupled mode theory. After some repackaging, I tried submitting it to a few different journals covering optics/photonics and astronomical instrumentation, ultimately unsuccessfully. In each case, the reason for the rejection was that the contents of the paper were out of scope and/or inaccessible to the journal audience. As such, I have opted to just post the paper on arXiv, in the case that the methods presented here are useful to others. My hope is that the manuscript at least highlights some cross-disciplinary ideas that could accelerate the development of astrophotonic devices, while providing outlines for the underlying mathematical framework at a lower level of rigor.
\\\\
To improve the accessibility of this version of the paper, I have added more context in the introduction sections, including a primer for the CMT which considers a mechanical analog. I have also included clarifications on some notational aspects (e.g. bra-ket notation), and added in some footnotes covering extra mathematical context and derivation details. Furthermore, I emphasize that it is not necessary to have an understanding of every topic mentioned in this work; the paper mentions concepts such as Hilbert spaces, differential forms, and fiber bundles more to make a connection that might be appreciated by readers who happen to have encountered those concepts before. If you have any suggestions (or comments, or questions, or if you catch a typo), please contact me. My email is at the top of the paper.
\section*{Acknowledgments}
I'd like to thank the anonymous referee from PASP, who provided several helpful suggestions which I eventually incorporated into the manuscript. I'd also like to thank my advisor, Mike Fitzgerald, for helpful discussions about the paper, and for sanity checking some of the math.

\bibliography{refs}

\end{document}